\documentclass[11pt]{article}
\usepackage{moriond,epsfig}
\usepackage{relsize}

\def\Journal#1#2#3#4{{#1} {\bf #2}, #3 (#4)}

\def\PLB{{\em Phys. Lett.}  B}
\def\PRL{\em Phys. Rev. Lett.}
\def\PRD{{\em Phys. Rev.} D}

\def\be{\begin{equation}}
\def\ee{\end{equation}}
\def\bea{\begin{eqnarray}}
\def\eea{\end{eqnarray}}

\begin{document}
\vspace*{4cm}
\title{Bottomonium Results from
 {\slshape B\kern-0.1em{\smaller A}\kern-0.1emB\kern-0.1em{\smaller A\kern-0.2em R}}  and BELLE}

\author{J\"org Marks for the {\slshape B\kern-0.1em{\smaller A}\kern-0.1emB\kern-0.1em{\smaller A\kern-0.2em R}} Collaboration }

\address{Physikalisches Institut der Universit\"at Heidelberg \\
Philosophenweg 12,
D-69120 Heidelberg, GERMANY}

\maketitle\abstracts{
After nine years of operation the 
{\slshape B\kern-0.1em{\smaller A}\kern-0.1emB\kern-0.1em{\smaller A\kern-0.2em R}} 
experiment at the B factory PEPII
(Standford Linear Accelerator Center) stopped data taking in April 2008.
The last three month of data taking were devoted to $e^+e^-$ collisions at
center of mass energies of the $\Upsilon(2S)$, $\Upsilon(3S)$ and to 
an energy scan above the $\Upsilon(4S)$. Besides the observation of 
the bottomonium ground state $\eta_b$,
the center of mass energy dependent $e^+e^- \rightarrow b\bar{b}$ cross section was measured in the energy range from 10.54 to 11.20 GeV.
BELLE observed an enhancement in the  production cross section
for $e^+ e^- \rightarrow \Upsilon(nS) \pi^+ \pi^- \rightarrow  \mu^+ \mu^- \pi^+ \pi^-$ in an energy scan from 10.83 to 11.02 GeV.}
\section{Introduction}
The bound states of $b\bar{b}$, the bottomonium states, are the
heaviest and most compact bound states of quarks
and anti quarks in nature.  They were first discovered
as spin triplet states called $\Upsilon$ by the E288
collaboration at Fermilab in 1977 in $p$ scattering on 
Cu and Pb targets studying muon pairs in a regime
of invariant masses larger than 5 GeV~\cite{UpsilonDiscovery}. 
Thirty years after the discovery
of these $b \bar{b}$ triplet states, still no evidence for 
the lowest energy spin singlet state, the pseudo scalar
$\eta_b$, was found.

Spectroscopic measurements of fine and hyperfine structure splittings 
of hadronic and radiative transitions
in the bottomonium system allow to test calculations of
 NRQCD, QCDME and lattice QCD. In particular, the hyperfine
mass splitting between the singlet and triplet states yields information
about the spin-spin interactions. Of the recent topics in bottomonium physics,
{\slshape B\kern-0.1em{\smaller A}\kern-0.1emB\kern-0.1em{\smaller A\kern-0.2em R}}'s 
discovery of the $\eta_b$
and the measurement of the hyperfine splitting are 
discussed~\cite{EtabDiscovery}. Results of an inclusive $b \bar{b}$ cross
section measurement of a precision energy scan above 
the $\Upsilon (4S)$ are presented. These results are compared to an
exclusive cross section measurement 
of $e^+e^- \rightarrow \Upsilon(nS) \pi^+ \pi^-$  by 
BELLE in a scan on the $\Upsilon(5S)$
resonance.
\section{Discovery of the $\eta_b$ Meson}
The large BABAR dataset on $\Upsilon (3S)/(2S)$ of 120 million/100 million events allows to search for the rare radiative
M1 transitions from the triplet states $\Upsilon(3S)$ and $\Upsilon(2S)$ to the $\eta_b$. 

The strategy is to search
in the inclusive photon spectrum for the
decay $\Upsilon(3S) \rightarrow \gamma \eta_b$ in the center of mass
frame of the $\Upsilon(3S)$. Besides the signal photons at an energy of about 900 MeV, we expect large backgrounds of non-peaking 
and peaking nature.
Continuum $q\bar{q}$ events, $\Upsilon(3S)$ cascade
decays and $\Upsilon(3S) \rightarrow \gamma g g$ events contribute
to the non-peaking background.
There are two contributions to the peaking background: i) the decay chain from $\Upsilon(3S)$ to the triplet\footnote{The 3 states of the $\chi_{bJ}(2S)$ decaying to $\Upsilon(1S)$ appear as one peak at about 760 MeV due to energy 
resolution and Doppler broadening.} $\chi_{bJ}$, which then decays to $\Upsilon(1S)$, ii) initial state 
radiation (ISR) with a photon of such a radiated
energy ($E_{\gamma} \approx$ 860 MeV) that the remaining virtual photon matches the $\Upsilon(1S)$.

Knowing all sources entering the inclusive photon spectrum, for each contribution a probability density
function (PDF) is determined. 
A binned maximum likelihood fit in the photon energy range from 500 to 1100 MeV
allows to extract the $\eta_b$ signal.  About 10~\% of the data are used to improve the PDF determination, the event selection and the background
suppression. This data are
discarded in the final analysis.
The shape of the photon distribution ($E_{\gamma} = s -m^2 / 2 \sqrt{s}$) of the decay to $\eta_b$ is determined from 
MC as a convolution of a Crystal Ball and a Breit-Wigner function. 
The width of the Breit-Wigner function is fixed to 10 MeV and variations are considered as systematic errors.
For the non-peaking background component an exponential Ansatz is used; the starting parameters 
are determined from the side bands. The $\chi_{bJ}(2S)$ decays are parametrized as 3 Crystal Ball functions. 
Their width is fixed and is for all 3 lines the same. The relative peak positions are taken from PDG.
The relative yields are also fixed.
In the final fit the yield of the contribution from ISR ($e^+e^- \rightarrow \gamma_{ISR} \Upsilon(1S)$) 
 is fixed and taken from the extrapolated yield of the $\Upsilon(4S)$
off-peak data to the $\Upsilon(3S)$ on-peak sample taking
the luminosity, the reconstruction efficiency and the cross section into account.
\begin{figure}[htb]
\begin{center}
\epsfig{file=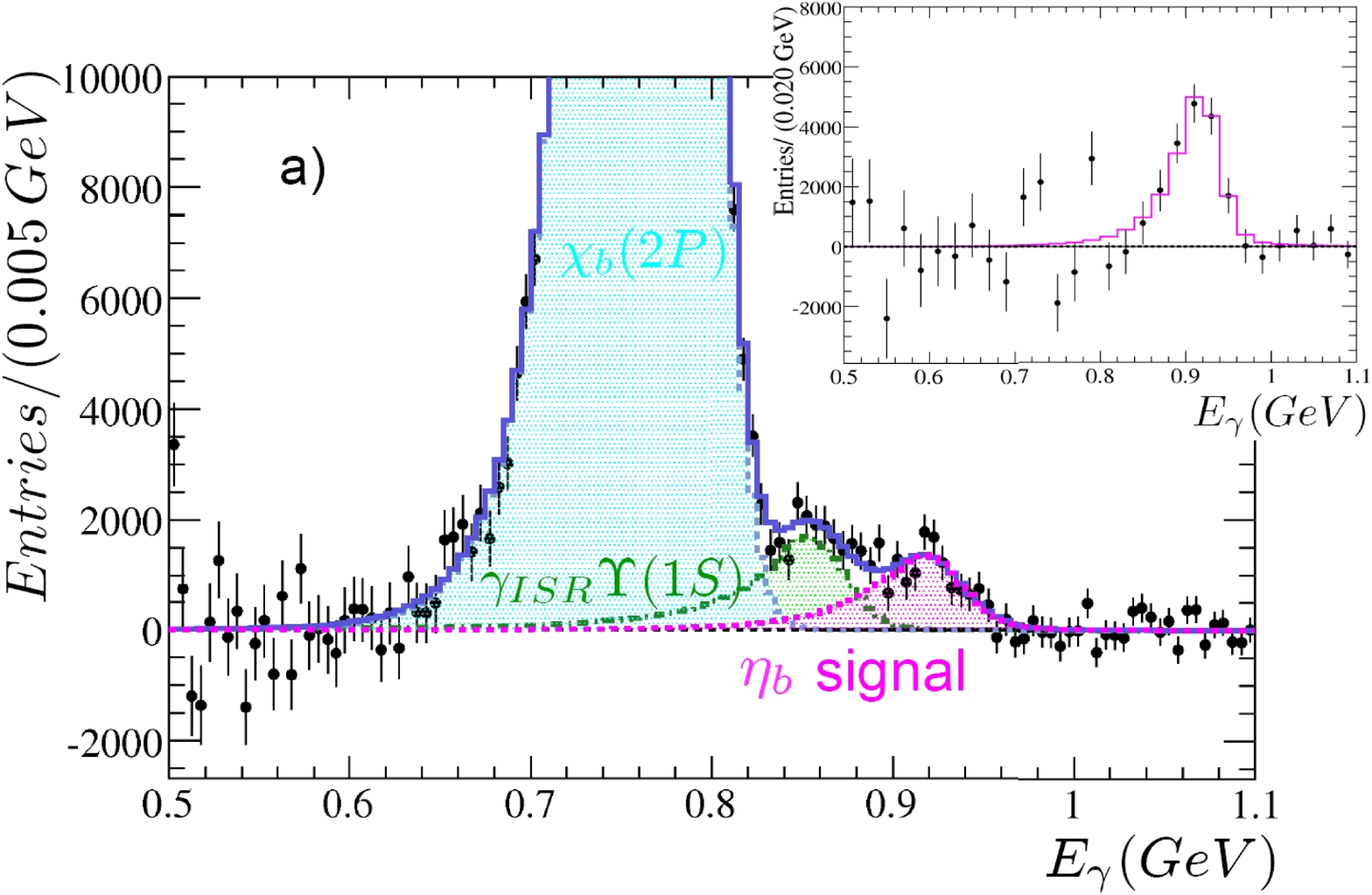,height=1.5in}
\hspace{1.0in}
\epsfig{file=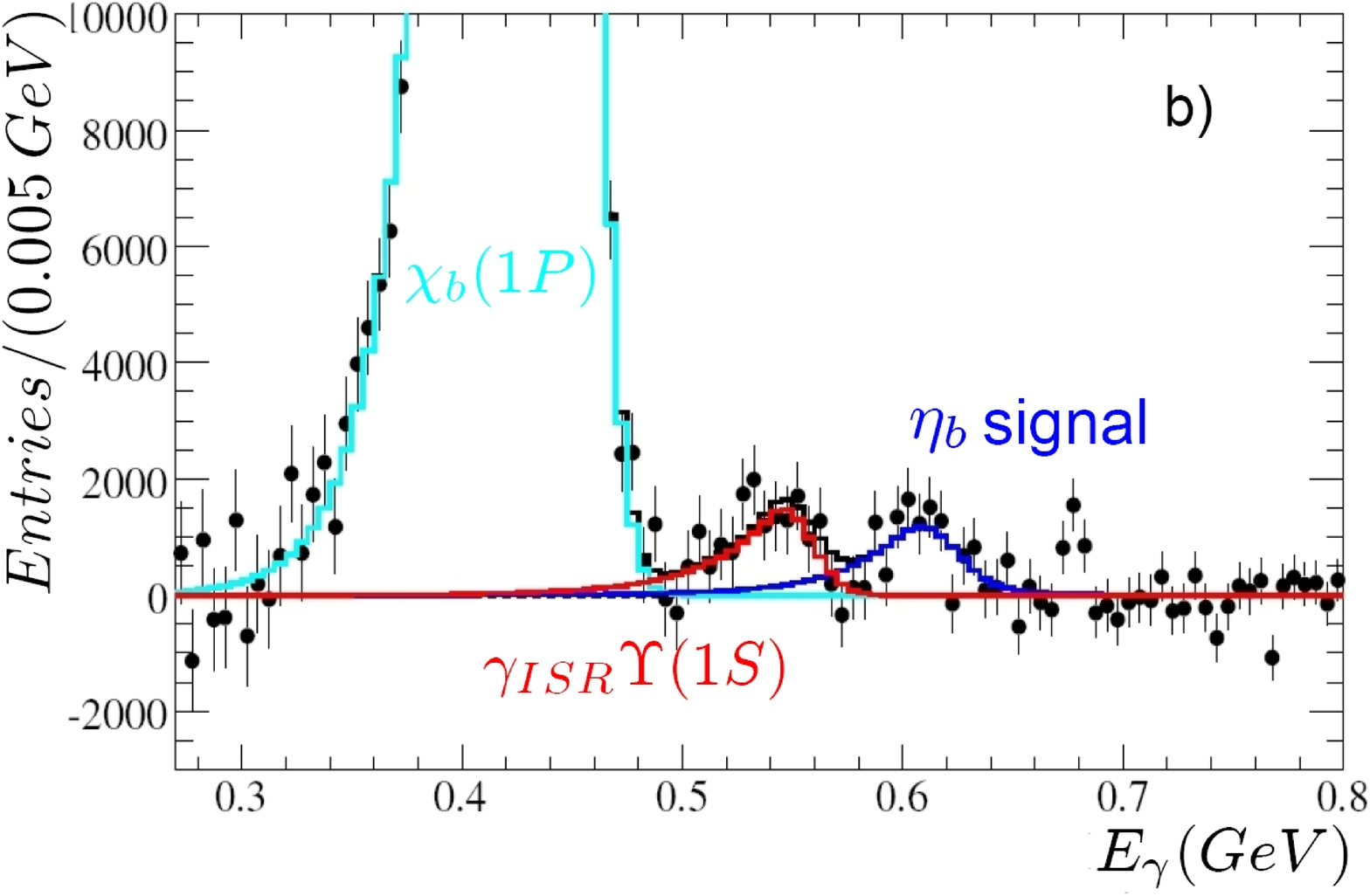,height=1.5in}
\end{center}
\caption{\label{eta_bPlot} Inclusive photon spectrum after subtracting the non-peaking background. a)
For $\Upsilon(3S)$ data with PDF's for $\chi_{bJ}(2S)$ peak (cyan), ISR $\Upsilon(1S)$ (green), $\eta_b$ signal (magenta)
and the sum of all three. The inline plot is the inclusive photon spectrum after subtracting 
all components except the $\eta_b$ signal. b) For $\Upsilon(2S)$ data with PDF's for $\chi_{bJ}(1S)$ peak (cyan), 
ISR $\Upsilon(1S)$ (red) and $\eta_b$ signal (blue).}
\end{figure}

A maximum likelihood fit of the four components to the data sample with an integrated luminosity of 25.6 $fb^{-1}$ 
(109 million $\Upsilon(3S)$ events) is performed. Figure~\ref{eta_bPlot}a shows 
the inclusive photon spectrum and the PDFs of the fit result
as colored lines after subtracting
the non-peaking background. The $\chi_{bJ}(2S)$ contribution is indicated in light blue, in green the contribution
from initial state radiation. The $\eta_b$ peak in magenta is clearly visible. Subtracting the $\chi_b$ 
and ISR contributions leads to the $\eta_b$ signal shown in the upper right part of Figure~\ref{eta_bPlot}a. The
photon energy is measured to be $\langle E_{\gamma} \rangle = 921.2^{+2.1}_{ -2.8} \pm 2.4 \;  MeV$ with  a significance
of $10 \sigma$.

In addition to the $\eta_b$ search in $\Upsilon (3S)$ data,
{\slshape B\kern-0.1em{\smaller A}\kern-0.1emB\kern-0.1em{\smaller A\kern-0.2em R}}
performed a similar analysis using 92 Million $\Upsilon(2S)$ events~\cite{Etabin2sDiscovery}.
The $\eta_b$ discovery is confirmed in this channel with a signal significance of $3.5 \sigma$.  
Both values of the $\eta_b$ mass agree very well. The combined mass of the $\eta_b$ is measured to be 
$M_{\eta_{b}} = 9390.4 \pm 3.1 \; MeV/c^2$, which is in good agreement with unquenched lattice QCD 
calculations~\cite{metablattice}. Using the PDG average for the mass of the $\Upsilon(1S)$, 
{\slshape B\kern-0.1em{\smaller A}\kern-0.1emB\kern-0.1em{\smaller A\kern-0.2em R}} 
measures a hyperfine mass splitting
of $\Delta M_{\Upsilon(1S) - \eta_{b}} = 69.9  \pm 3.1 \; MeV/c^2$ well in agreement with lattice QCD
predictions~\cite{deltamlattice}. 
The ratio of the branching fraction measurements for $\Upsilon(3S) \rightarrow \eta_b \gamma$ and 
$\Upsilon(2S) \rightarrow \eta_b \gamma$ is $ R_{{\mathcal{B}}} = \mathcal{B} (\Upsilon(2S) \rightarrow  \gamma \eta_b) / \mathcal{B} (\Upsilon(3S) \rightarrow  \gamma \eta_b)  =  0.89 ^{+0.25+0.12}_{-0.23-0.16} $. According 
to Godfrey and Rosner~\cite{BRM1}, this is compatible with the assumption of radiative M1 transitions.    

\section{Energy Scans above $\Upsilon(4S)$}
Recently, non-baryonic charmonium states which do not behave like standard $c\bar{c}$ states were discovered.
The question arises, if similar exotic states with $J^{PC}=1^{--}$ appear
in the bottomonium energy regime. Scaling the
Y states (4260, 4350, 4660)  from the charmonium to the bottomonium regime, the interesting energy 
range is above $\Upsilon(4S)$ and below 11.2 GeV.
{\slshape B\kern-0.1em{\smaller A}\kern-0.1emB\kern-0.1em{\smaller A\kern-0.2em R}} performed a scan 
in the center of mass energy from 10.54 to 11.2 GeV in 5 MeV steps with 25 $pb^{-1} $ of recorded data per point.
This is about 4 times finer with a 30 times larger amount of data than the last scan done 25 years ago 
at CESR~\cite{EscanCLEO,EscanCUSP}. Including 8 additional points of irregular spacing 
on $\Upsilon(6S)$, the total amount of data corresponds to an integrated luminosity of 3.9 $fb^{-1}$.

{\slshape B\kern-0.1em{\smaller A}\kern-0.1emB\kern-0.1em{\smaller A\kern-0.2em R}} 
follows an inclusive approach to search 
for new states with $b$ quark content measuring the inclusive hadronic cross section 
as the ratio $R_b (s) =  \sigma_{bb ( \gamma ) }(s) / \sigma^{0}_{\mu\mu} (s)$ 
at different center of mass energies~\cite{Escanbabar}.
Here,  $\sigma_{bb ( \gamma ) }(s)$ is the total cross section of $e^+ e^- \rightarrow b \bar{b} (\gamma)$ including 
the $b\bar{b}$ states produced in initial state radiation below the open beauty threshold and 
$ \sigma^{0}_{\mu\mu} (s)$ is the lowest order cross section of $e^+ e^- \rightarrow \mu^+ \mu^-$.
\begin{figure}[htb]
\begin{center}
\epsfig{file=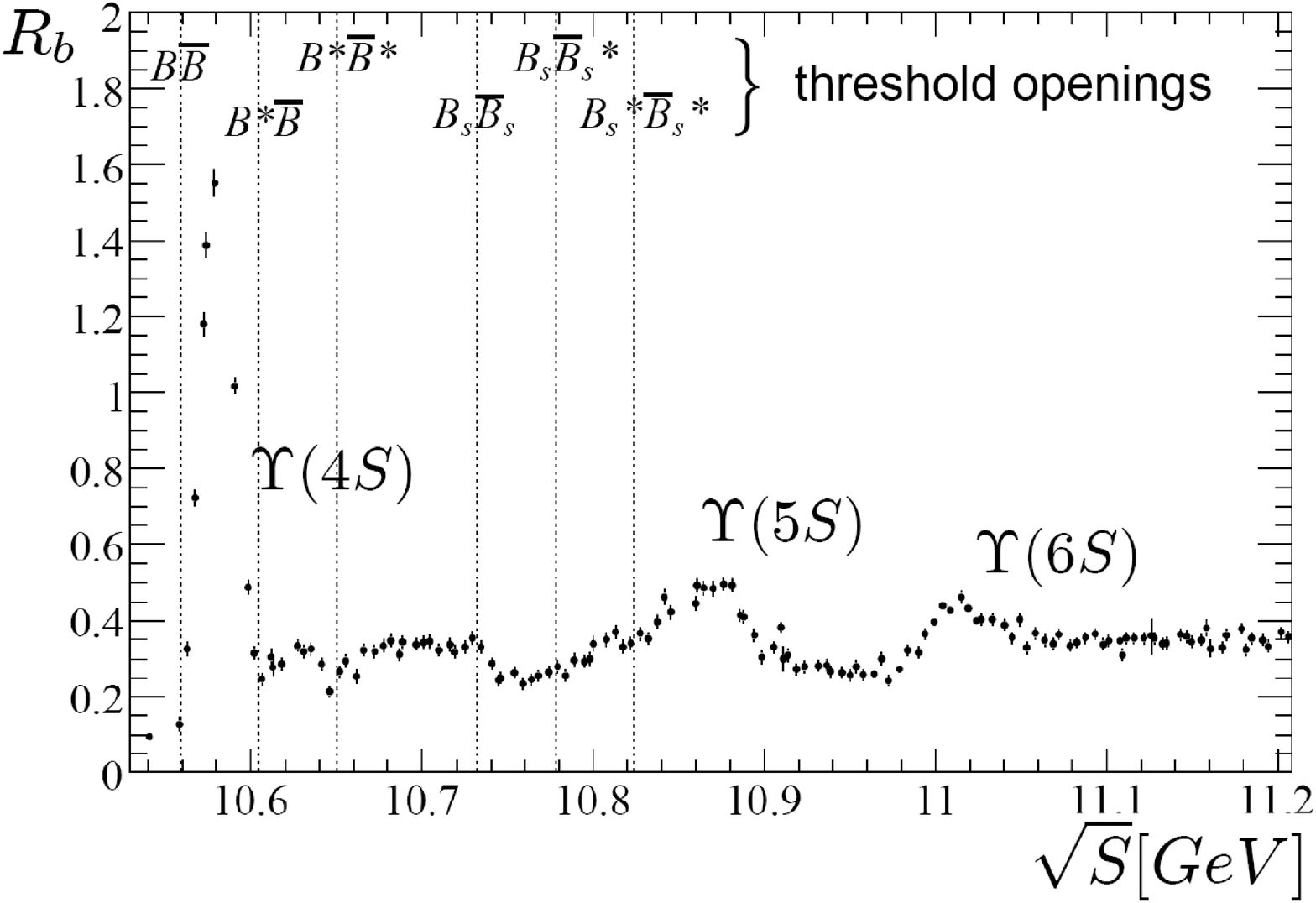,height=1.4in}
\hspace{1.0in}
\epsfig{file=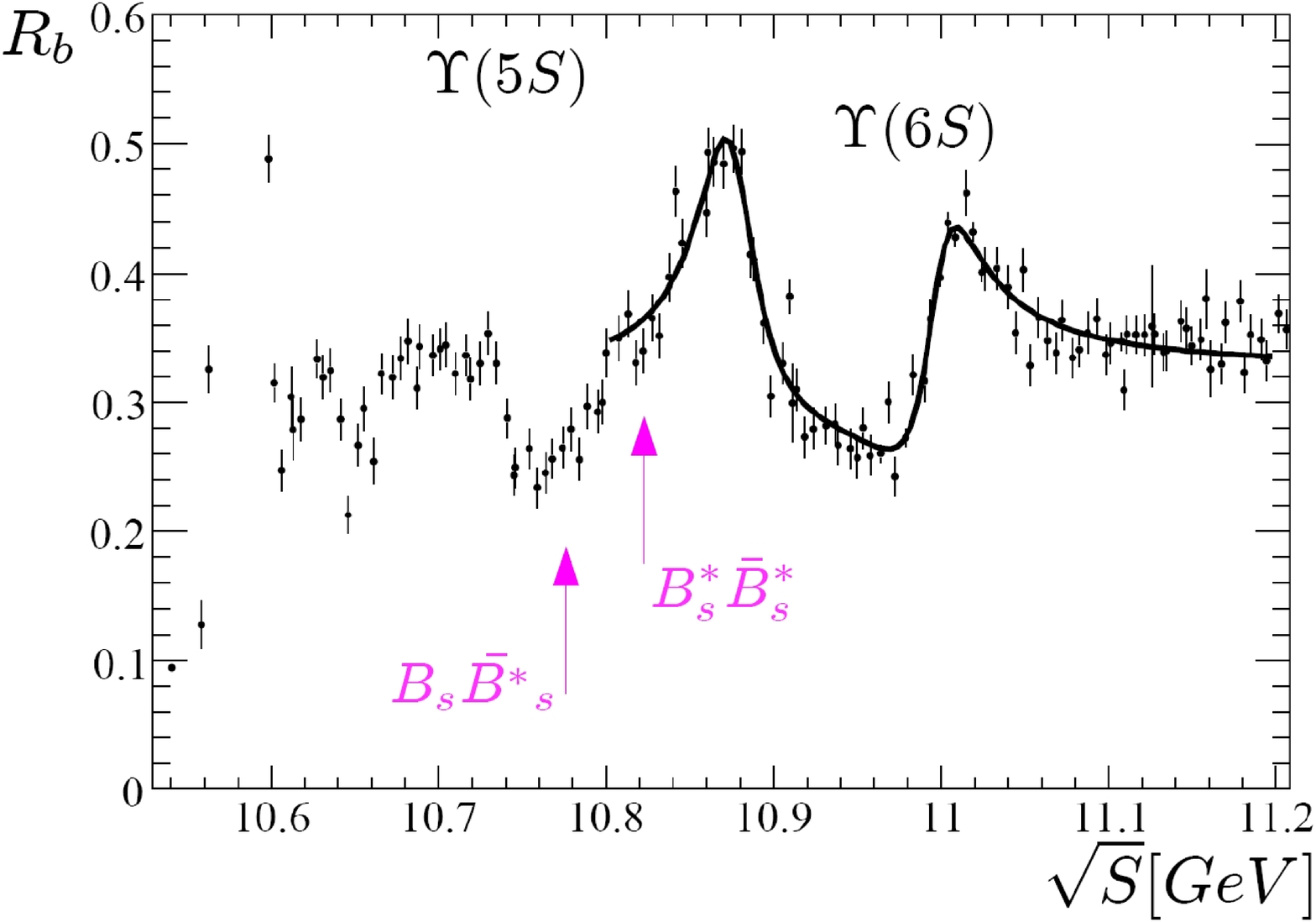,height=1.4in}
\end{center}
\caption{\label{escanPlot}(Left)$R_b$ as function of the center of mass energy with the position of the opening
thresholds of the $e^+ e^- \rightarrow B_{(s)}^{(*)}\bar{B}_{(s)}^{(*)}$ processes as dotted lines. 
(Right) Zoom of the same plot with the fit result superimposed.}
\end{figure}
The region above the $\Upsilon(4S)$ is explored with unprecedented details as shown in Figure~\ref{escanPlot} 
by the measurement of $R_b$ as function of the center of mass energy. 
The errors are of statistical and uncorrelated systematic nature. 
The dotted lines indicate the different $B$ meson production thresholds. 
The large statistics per energy point and the small energy steps reveal structures
which seem to correspond to threshold openings. 
The $\Upsilon(5S)$ and $\Upsilon(6S)$  candidates are probably not pure resonance structures 
as predicted  within the coupled channel 
model in 1984  by T\"ornquist~\cite{Escantoernquist}. It handles the coupling between the quarkonia and the continuum. 
Coupled channel effects play a significant role in accounting 
for the energy spacing of the $nS$ level. All resonances contribute by interference with 
the dominant resonance. Therefore, an interpretation of the measured structures is very difficult.
The bumps in the region from 10.6  to 10.75 GeV are not due to resonances, 
but appear due to threshold openings of the $B^*\bar{B}$ and $B^*\bar{B^*}$ 
and 
the node structure in the overlap integrals. Above $\Upsilon(6S)$ a plateau is clearly visible.

In order to determine the parameters for the $\Upsilon(5S)$ and $\Upsilon(6S)$ candidates, the following simplified model is fit to the data in the energy range from 10.8 to 11.2 GeV:
$\sigma = |A_{nr}|^2 + |B_{r}  + A_{5S} e^{i\phi_{5S}} BW(M_{5S},\Gamma_{5S})  + A_{6S} e^{i\phi_{6S}} BW(M_{6S},\Gamma_{6S}) |^2$, 
$BW(M,\Gamma)$ is a relativistic Breit-Wigner resonance.  The values obtained $M(\Upsilon(5S)) = 10876 \pm 2  \; MeV/c^2 $, 
$\Gamma (\Upsilon(5S)) = 43 \pm 4  \; MeV/c^2$  and 
 $M(\Upsilon(6S)) = 10960 \pm 2  \; MeV/c^2$, $\Gamma (\Upsilon(6S)) = 37 \pm 3 \;  MeV/c^2$ differ significantly from
the PDG values  $M(\Upsilon(5S) = 10865 \pm 8 \; MeV/c^2 $, 
$\Gamma (\Upsilon(5S) = 110 \pm 13  \;MeV/c^2$  and 
$M(\Upsilon(6S)) = 11019 \pm 8  \; MeV/c^2$, $\Gamma (\Upsilon(6S)) = 79 \pm 16  \; MeV/c^2$. The result of the fit is superimposed
in Figure~\ref{escanPlot} (right). 
The number of states and their energy dependence is a priori unknown. Therefore, a calculation
within a proper coupled channel approach would certainly yield different results.

\begin{figure}[htb]
\begin{center}
\epsfig{file=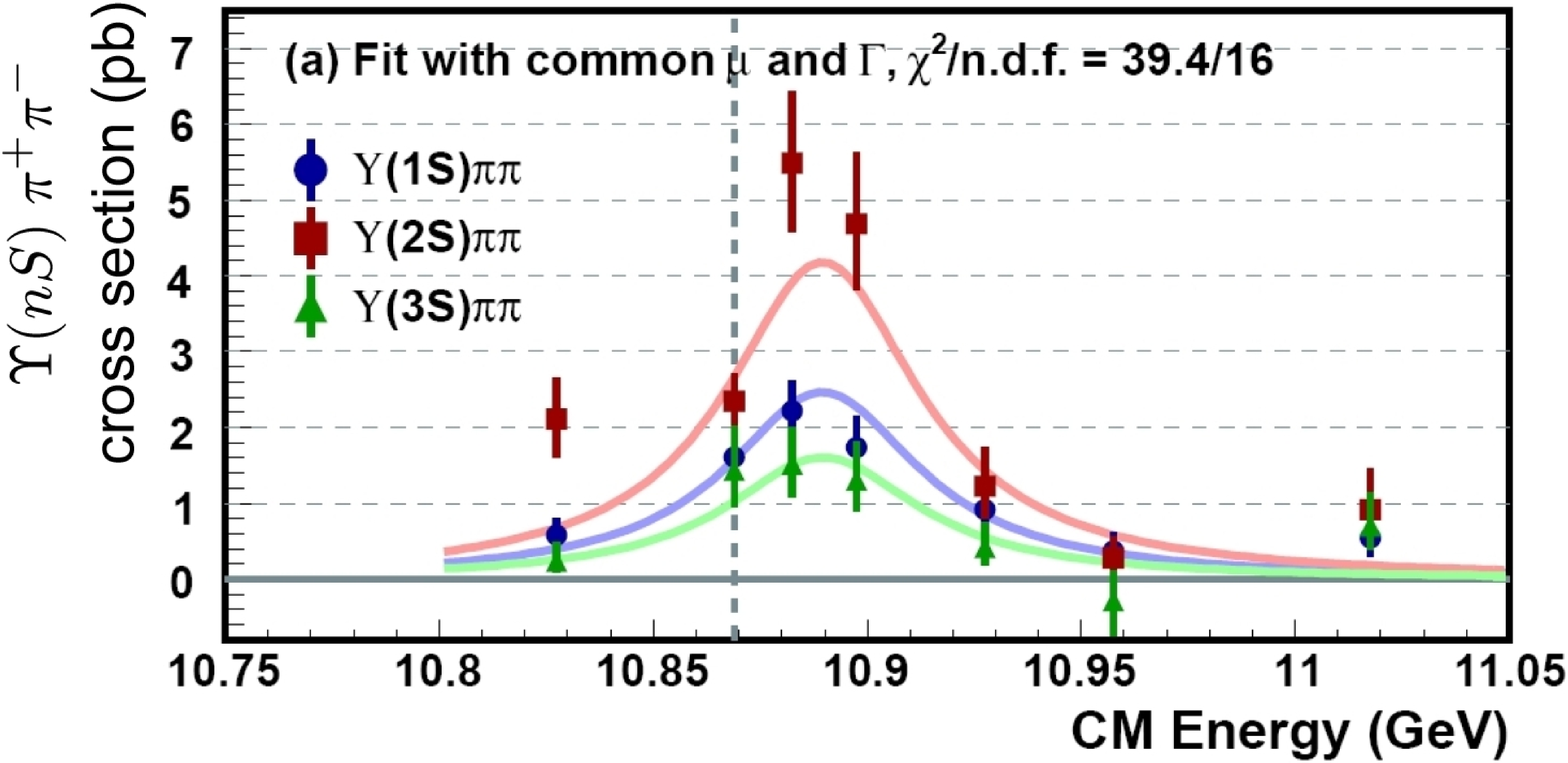,height=1.3in}
\hspace{0.5in}
\epsfig{file=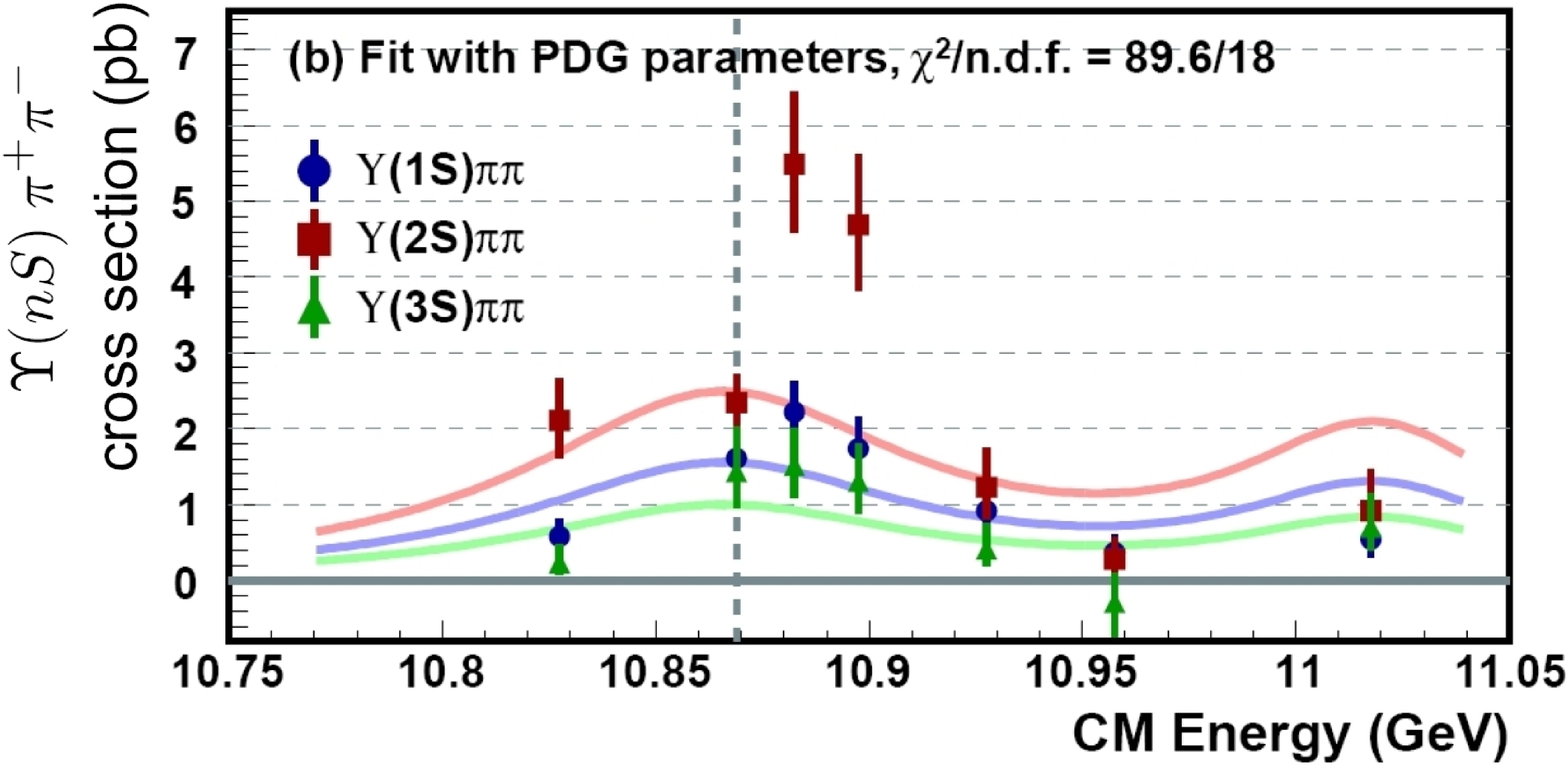,height=1.3in}
\end{center}
\caption{\label{escanBellePlot} The center of mass energy dependent cross sections for 
$e^+ e^- \rightarrow \Upsilon(nS) \pi^+ \pi^- \rightarrow  \mu^+ \mu^- \pi^+ \pi^-$ ($n=1,2,3$) processes.
a) The results of a fit with a common mean and width of an S-wave Breit-Wigner model are shown as curves.
b) The results of a fit with the PDG $\Upsilon(5S)$ and  $\Upsilon(6S)$ parameters is superimposed.}
\end{figure}
In contrast to 
{\slshape B\kern-0.1em{\smaller A}\kern-0.1emB\kern-0.1em{\smaller A\kern-0.2em R}},
BELLE followed an exclusive approach measuring the energy dependence of the cross section 
of $e^+ e^- \rightarrow \Upsilon(nS) \pi^+ \pi^- \rightarrow  \mu^+ \mu^- \pi^+ \pi^-$ ($n=1,2,3$) in
an energy scan within the $\Upsilon(5S)$ region~\cite{Escanbelle2,Escanbelle1}.
Data of six energy points from 10.83 to 11.02 GeV corresponding to an integrated 
luminosity of 7.9 $fb^{-1}$ were collected.
The signal yield for the cross section measurement is extracted by an unbinned maximum likelihood fit
to $\Delta M$, defined as the difference between 
$M(\mu^+ \mu^- \pi^+ \pi^-)$ and $M(\mu^+ \mu^-)$, for the 3 different resonance regions
$\Upsilon(1S)$, $\Upsilon(2S)$ and $\Upsilon(3S)$.  
The three sets of cross section measurements as a function of the center of mass energy are shown in Figure~\ref{escanBellePlot} in
different colors. The fit of  
a single S-wave Breit-Wigner resonance model is superimposed (Figure~\ref{escanBellePlot}a).
In the fit the normalization as well as a common mean $\mu(\Upsilon(5S))$ and width $\Gamma(\Upsilon(5S))$ 
are extracted. An enhancement in the production of the final states is observed and the 
conventional $\Upsilon(5S)$ lineshape 
does not describe the measurements well.
The values obtained in the fit, $\mu(\Upsilon(5S)) = 10889.6 \pm 1.8 \pm 1.9 \;  MeV/c^2$ and 
$\Gamma(\Upsilon(5S)) = 54.7^{+8.5}_{-7.2} \pm 2.5 \; MeV/c^2$, are clearly different from the PDG values 
listed above. This is supported in
Figure~\ref{escanBellePlot}b where a fit with the PDG  $\Upsilon(5S)$ and  $\Upsilon(6S)$ parameters yields a poor
$\chi^2$ value; the observed resonance structure disagrees with the  
$\Upsilon$ states given by the PDG.  
\section{Summary}
The large {\slshape B\kern-0.1em{\smaller A}\kern-0.1emB\kern-0.1em{\smaller A\kern-0.2em R}} datasets on $\Upsilon(2S) / \Upsilon(3S)$ resulted in the discovery of the lowest energy spin
singlet state of the bottomonium system $\eta_b$ in $\Upsilon(3S) \rightarrow \eta_b \gamma$ decays. The $\eta_b$ mass was measured to be $M_{\eta_{b}} = 9390.4 \pm 3.1 \; MeV/c^2$ with a hyperfine splitting $\Delta M_{\Upsilon(1S) - \eta_{b}} = 69.9  \pm 3.1 \; MeV/c^2$.     

These measurements were complemented by an inclusive hadronic cross section measurement above $\Upsilon(4S)$ from 10.54 to 11.2 GeV which revealed structures with unprecedented detail. {\slshape B\kern-0.1em{\smaller A}\kern-0.1emB\kern-0.1em{\smaller A\kern-0.2em R}} extracted from the fit of a simplified model parameters for $\Upsilon(5S)$ and  $\Upsilon(6S)$ which indicate a smaller width than the PDG values. This is supported by a cross section measurement from BELLE of $e^+ e^- \rightarrow \Upsilon(nS) \pi^+ \pi^- \rightarrow  \mu^+ \mu^- \pi^+ \pi^-$ in the  $\Upsilon(5S)$ region of 10.83 to 11.02 GeV.

\end{document}